\documentclass[aps,prl,twocolumn,superscriptaddress,showpacs,longbibliography]{revtex4-2}

\usepackage{CJK}
\usepackage{graphicx}
\usepackage{bm}
\usepackage{amssymb}
\usepackage{amsmath}
\usepackage[colorlinks,linkcolor=blue,anchorcolor=blue,citecolor=blue]{hyperref}
\usepackage{xcolor}
\usepackage{graphicx} 
\usepackage{xcolor}
\usepackage{textcomp}
\usepackage{lettrine}
\usepackage{ragged2e}

\usepackage{subcaption}
\usepackage{comment}
\DeclareUnicodeCharacter{2218}{\ensuremath{\circ}}
\usepackage{xr}
\usepackage{caption}
\usepackage{newunicodechar}
\newunicodechar{₂}{$_2$}
\begin{document}

\title{Coexistence and tunability of orbital and spin Hall effects in RuO\textsubscript{2}}

\author{Lishu Zhang}
\email{lishu.zhang@sdu.edu.cn}
\affiliation{Key Laboratory for Liquid-Solid Structural Evolution and Processing of Materials, Ministry of Education, Shandong University, Jinan 250061, China}

\author{Mahmoud Zeer}
\affiliation{Peter Gr{\"u}nberg Institut (PGI-1) and Institute for Advanced Simulation (IAS-1), Forschungszentrum J{\"u}lich, J{\"u}lich 52428, Germany}
\affiliation{Department of Physics, RWTH Aachen University, 52056 Aachen, Germany}

\author{Dongwook Go}
\affiliation{Peter Gr{\"u}nberg Institut (PGI-1) and Institute for Advanced Simulation (IAS-1), Forschungszentrum J{\"u}lich, J{\"u}lich 52428, Germany}
\affiliation{Institute of Physics, Johannes Gutenberg University Mainz, 55099 Mainz, Germany}

\author{Theodoros Adamantopoulos}
\affiliation{Peter Gr{\"u}nberg Institut (PGI-1) and Institute for Advanced Simulation (IAS-1), Forschungszentrum J{\"u}lich, J{\"u}lich 52428, Germany}
\affiliation{Department of Physics, RWTH Aachen University, 52056 Aachen, Germany}

\author{Peter Schmitz}
\affiliation{Peter Gr{\"u}nberg Institut (PGI-1) and Institute for Advanced Simulation (IAS-1), Forschungszentrum J{\"u}lich, J{\"u}lich 52428, Germany}
\affiliation{Department of Physics, RWTH Aachen University, 52056 Aachen, Germany}

\author{Stefan Blügel}
\affiliation{Peter Gr{\"u}nberg Institut (PGI-1) and Institute for Advanced Simulation (IAS-1), Forschungszentrum J{\"u}lich, J{\"u}lich 52428, Germany}

\author{Chengwang Niu}
\affiliation{School of Physics, Shandong University, Jinan 250100, China}

\author{Yuriy Mokrousov}
\affiliation{Peter Gr{\"u}nberg Institut (PGI-1) and Institute for Advanced Simulation (IAS-1), Forschungszentrum J{\"u}lich, J{\"u}lich 52428, Germany}
\affiliation{Institute of Physics, Johannes Gutenberg University Mainz, 55099 Mainz, Germany}

\author{Shishen Yan}
\email{shishenyan@sdu.edu.cn}
\affiliation{School of Physics, Shandong University, Jinan 250100, China}

\author{Hyunsoo Yang}
\affiliation{Department of Electrical and Computer Engineering, National University of Singapore, 4 Engineering Drive 3, Singapore, 117583, Republic of Singapore}

\author{Lei Shen}
\email{shenlei@nus.edu.sg}
\affiliation{Department of Mechanical Engineering, National University of Singapore, 9 Engineering Drive 1, Singapore, 117575, Republic of Singapore}

\begin{abstract}
Altermagnetic materials, especially RuO$_2$, have recently attracted considerable attention for their unique magnetic properties and energy-efficient spintronic applications. However, recent experimental studies have reported highly conflicting signatures regarding altermagnetic spin splitting and charge--spin interconversion (CSI) in RuO$_2$. While some experiments link efficient CSI to non-relativistic altermagnetic spin-splitting effects, others observe large CSI signals in non-spin-splitting RuO$_2$, which are instead explained by relativistic inverse spin Hall effects. In this work, based on first-principles calculations, we reveal that these controversial experimental results originate from a phase-dependent coexistence and relative dominance of the orbital Hall effect (OHE) and spin Hall effect (SHE) in RuO$_2$. We systematically investigate the OHE and SHE in both altermagnetic and nonmagnetic phases of RuO$_2$. Our results show that the altermagnetic state hosts a giant OHE that exceeds the SHE by two orders of magnitude and carries an opposite sign. This dominant OHE can generate experimentally observed ``SHE-like'' voltages through orbital-to-spin conversion, explaining previously reported altermagnetic CSI signals. In contrast, OHE of nonmagnetic RuO$_2$ is suppressed and a large relativistic SHE emerges, in agreement with recent angle-resolved photoemission and spin-pumping experiments. Finally, we demonstrate that the coexistence of OHE and SHE is tunable via chemical doping, enabling on-demand modulation of CSI in in RuO$_2$. Our work provides a new physical mechanism for understanding CSI in RuO$_2$ and highlights the central role of orbital transport.
\end{abstract}

\maketitle


Recently, a novel class of collinear magnetic materials characterized by the coexistence of crystal rotational symmetries and broken spatial inversion--time-reversal ($\mathcal{PT}$) symmetry, termed \emph{altermagnets}, has attracted extensive theoretical and experimental interest \cite{zhou2025manipulation, PhysRevX.12.040501, https://doi.org/10.1002/adfm.202409327,  tamang2025altermagnetism, gu2025ferroelectric, duan2025antiferroelectric, ding2025ferroelastically}. Distinct from conventional ferromagnets and antiferromagnets, altermagnets can host large \emph{non-relativistic} spin splitting even in the absence of net magnetization or strong spin--orbit coupling (SOC) \cite{naka2019spin, krempasky2024altermagnetic, li2024spectroscopic, fender2025altermagnetism, yuan2020giant}. This unique property gives rise to unconventional magnetotransport phenomena \cite{han2025harnessing, zarzuela2025transport, fang2024quantum, he2024quasi, zhou2025piezomagnetism}. Despite rapid progress, the field of altermagnetism remains in its early stage, and several fundamental issues are still under active debate. In particular, while spin- and anomalous-Hall effects (SHE and AHE), which rely on spin angular momentum, have been widely discussed in altermagnets \cite{song2025altermagnets, sheoran2025spontaneous, chu2025third, xu2025identification}, the orbital Hall effect (OHE) associated with orbital angular momentum has received far less attention, both theoretically and experimentally \cite{go2021, cuono2023orbital, kontani2009giant, yahagi2024neel, go2018, dou2025anisotropic}. This gap is especially evident in RuO$_2$, which has recently emerged as a prototypical platform for exploring unconventional charge--spin interconversion (CSI). However, it exhibits strikingly controversial experimental signatures.

On the one hand, Bai \emph{et al.} reported that the non-relativistic altermagnetic (AM) spin-splitting effect (SSE) in RuO$_2$ enables efficient CSI \cite{bai2023efficient}. Subsequent studies further supported the coexistence of SSE- and SHE-driven CSI with comparable efficiencies but opposite signs \cite{liao2024separation}. On the other hand, opposite conclusions have been drawn from spin-pumping and spin-torque ferromagnetic resonance measurements, which revealed large inverse SHE signals even in RuO$_2$ samples lacking altermagnetic spin splitting \cite{wang2024inverse}. Most strikingly, recent high-resolution spin-resolved angle-resolved photoemission spectroscopy (ARPES) experiments reported no observable altermagnetic spin splitting, instead finding electronic structures fully consistent with a calculated nonmagnetic (NM) RuO$_2$ \cite{liu2024absence}. These results challenge the existence of altermagnetism in RuO$_2$ and its proposed role in spin-splitting-driven CSI in previous works \cite{bai2023efficient, liao2024separation}. This raises a central question: Why do RuO$_2$ samples exhibit qualitatively opposite CSI signatures from fundamentally different physical mechanisms that some consistent with non-relativistic SSE, while others are dominated by relativistic SHE?

\begin{figure*}
\centering
\includegraphics[width=\linewidth]{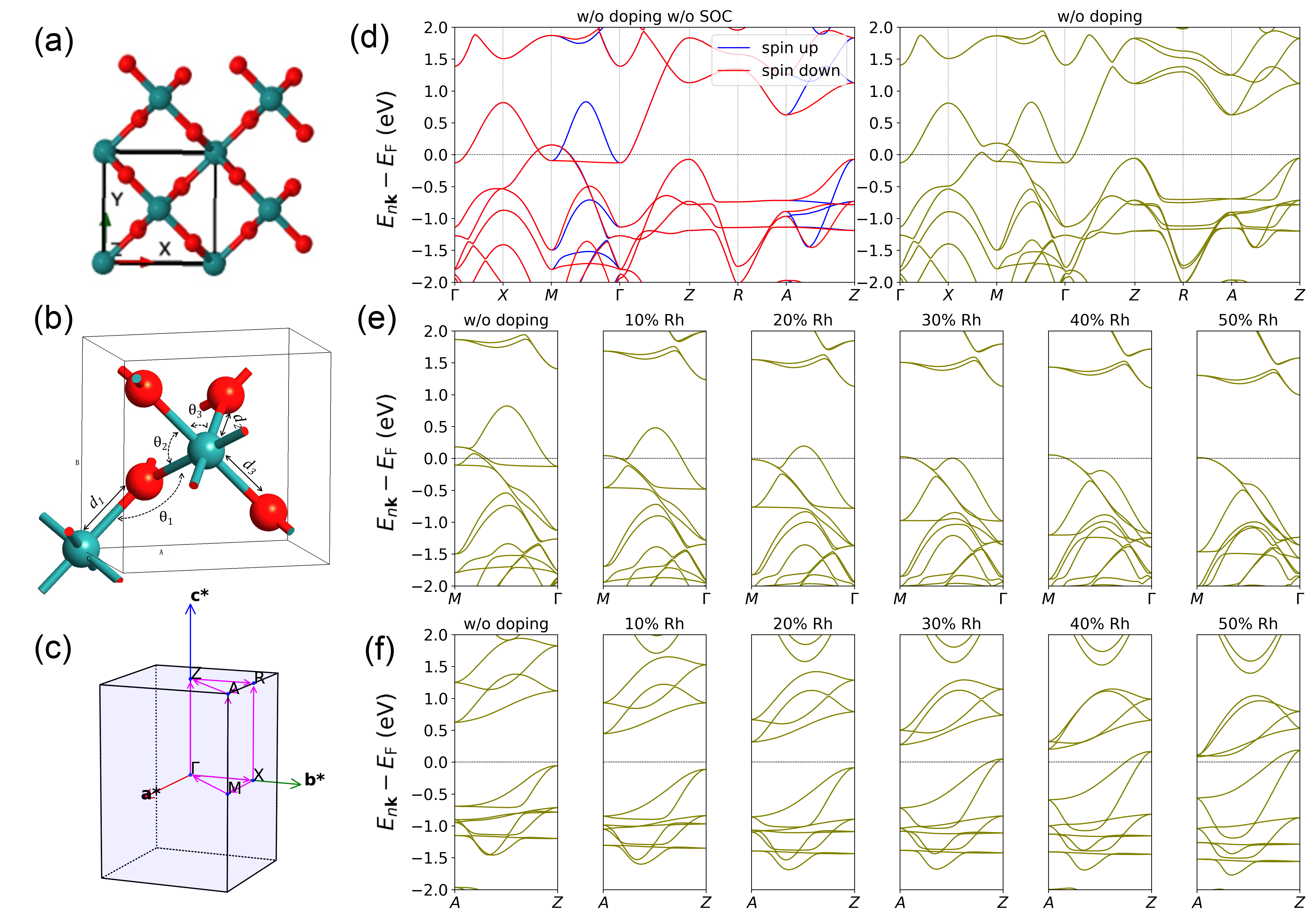}
\caption{
\justifying{
a) The crystal structure of RuO$_2$. b) and c) Schematic diagrams of charge-to-spin ($\sigma^{z}_{\mathrm{SHE}}$) and charge-to-orbital ($\sigma^{z}_{\mathrm{OHE}}$) current interconversion. d) and e) The calculated band structures of RuO$_2$ for the altermagnetic and non-magnetic state with SOC. f) and g) The calculated energy-dependent spin and orbital Hall conductivity of RuO$_2$ in b) and c) configurations. h) and i) Schematic explanation on the experimentally observed efficient CSI in AM and NM RuO$_2$, respectively. Through orbital-to-spin conversion, OHE dominates experimentally observed spin signals in AM RuO$_2$, while the emerging SHE becomes the main source in NM RuO$_2$.}}

\label{fig:1}
\end{figure*}

In this work, we clarify these seemingly contradictory experimental observations using first-principles calculations (see computational details in Supporting Information) for both altermagnetic and nonmagnetic RuO$_2$. We demonstrate that the AM and NM phases exhibit qualitatively distinct orbital and spin Hall responses. In the AM phase, non-relativistic band splitting changes the band structures and generates orbital Berry curvature hot spots, which gives rise to a giant orbital Hall effect that is two orders of magnitude larger than the spin Hall effect and has an opposite sign. Consequently, experimentally detected large spin currents can predominantly originate from orbital-to-spin conversion. In contrast, the NM phase suppresses the OHE while emerging a large relativistic SHE, which is consistent with SHE-dominated transport measurements and the absence of magnetic band splitting in ARPES experiments. Our results shed light on experimentally reported charge--spin interconversion in RuO$_2$, revealing that the coexistence and relative dominance of OHE and SHE govern the observed CSI behavior. Furthermore, we show that this coexistence can be efficiently tuned by chemical doping, enabling controlled generation of pure orbital currents, pure spin currents, as well as orbital and spin currents with either the same or opposite signs. These findings not only resolve a recent controversy surrounding altermagnetic RuO$_2$, but also highlight ``tunable" orbital transport as a key degree of freedom for future orbitronic device design.


Figures~\ref{fig:1}a-c show the atomistic crystal structure of RuO$_2$ and charge--spin(orbital) conversion configurations reported in the experiments Ref.~\cite{liao2024separation, wang2024inverse}. Ruthenium dioxide crystallizes in the rutile structure with space group $P4_2/mnm$, forming a tetragonal unit cell in which Ru atoms are octahedrally coordinated by six surrounding oxygen atoms (Fig.\ref{fig:1}a)\cite{https://doi.org/10.1002/adfm.202409327}. The critical Coulomb interaction strength of $U \approx 1.9$~eV (\textcolor{blue}{Table S1}) marks a transition from a nonmagnetic to an altermagnetic electronic state, as evidenced by the sudden onset of finite and opposite local moments on the two Ru sublattices (\textcolor{blue}{Table S1}). This decoupling between structural stability and magnetic instability provides an ideal platform for studying electronic and transport properties in both altermagentic and nonmagnetic RuO$_2$. In the AM phase, the two corner and central Ru atoms have identical but opposite spin, while all atoms carry zero magnetic moment in the NM phase. Figures~\ref{fig:1}d and e are the calculated band structures of altermagnetic and nonmagnetic RuO$_2$ in the presence of SOC. In the nonmagnetic phase, Kramer's degeneracy is preserved throughout the whole Brillouin zone, consistent with recent ARPES measurements \cite{liu2024absence}. 

Figures~\ref{fig:1}f and~\ref{fig:1}g show the calculated orbital and spin Hall conductivity of AM and NM RuO$_2$ under experimentally reported CSI configurations in Fig.~\ref{fig:1}b and c. As can be seen, at the Fermi level (highlighted by black circles), the orbital Hall response in the altermagnetic state is much larger than the spin Hall response in two orders of magnitude, while carrying an opposite sign. In contrast, the nonmagnetic state suppresses the orbital Hall effect and exhibits a sizable spin Hall conductivity. In AM RuO$_2$, the much larger OHE converts into experimentally observed ``SHE-like'' voltages (Fig.~\ref{fig:1}h), providing a natural explanation for previously reported altermagnetic CSI signals \cite{bai2023efficient}. In NM RuO$_2$ without spin splitting, the intrinsic large SHE dominates the observed CSI signals in the experiment \cite{wang2024inverse} due to the OHE conversion loss (Fig.~\ref{fig:1}i). Moreover, our OHE and SHE results also can explain the experimental observed an opposite sign for the CSI \cite{liao2024separation}.

\begin{figure}
\centering
\includegraphics[width=0.9\linewidth]{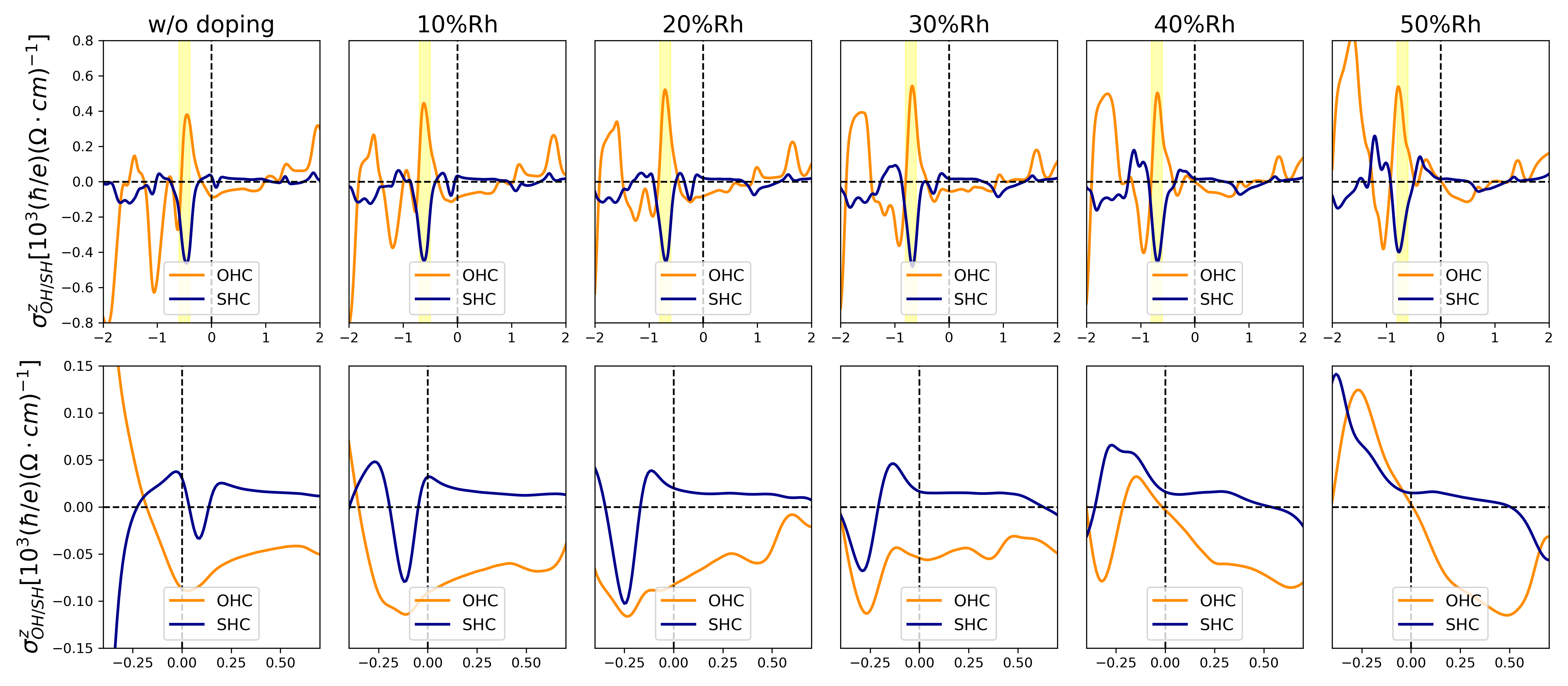}
\caption{
\justifying{
Momentum-resolved orbital and spin Berry curvatures in altermagnetic and non-magnetic RuO$_2$. Figure shows the $k$-resolved orbital and spin Berry curvatures ($\Omega^{z}{(\mathbf{k})}$) in the two-dimensional Brillouin zone ($k_{z}=0$) for altermagnetic (left column) and non-magnetic (right column) RuO$_2$. The Berry curvature hot spots and corresponding regions are highlighted by black circles.}}
\label{fig:2}
\end{figure}

To further understand the physics of OHE and SHE in altermagnetic and nonmagnetic RuO$_2$, we plot the momentum-space orbital and spin Berry curvature ($\Omega^{z}{(\mathbf{k})}$) into the distinct transverse transport mechanisms in AM and NM RuO$_2$ as shown in Fig.~\ref{fig:2}. As can be seen, the orbital Berry curvature in the altermagnetic state (left column) is strongly large and develops complex and anisotropic textures, whereas the spin Berry curvature is largely suppressed and close to zero. These orbital hot spots imply that the transverse response is primarily governed by orbital angular momentum rather than spin ones. In contrast, the nonmagnetic state (right column) displays a qualitatively different momentum-space responses. Its $k$-resolved spin Berry curvature exhibits pronounced and symmetry-related hot spots (highlighted by four black circles). This behavior is characteristic of a relativistic spin Hall effect and is consistent with experimental observations of dominant inverse spin Hall signals in nonmagnetic RuO$_2$ samples. The sharp contrast between the two magnetic phases indicates that altermagnetism profoundly reshapes the momentum-space distribution of Berry curvature and angular momentum transport. In particular, it explains why experimentally detected ``spin Hall-like'' voltages can arise from orbital currents via orbital-to-spin conversion in altermagnetic RuO$_2$, while truly spin-dominated Hall signals are recovered only in the nonmagnetic phase (Figs.\ref{fig:1}h and i). 

A complementary microscopic picture can be obtained from a minimal tight-binding model for the Ru $t_{2g}$ manifold as: 
\begin{equation}
\mathcal{H}(\mathbf{k})
=
\mathcal{H}_{\mathrm{hop}}(\mathbf{k})
+
\mathcal{H}_{\mathrm{cf}}
+
\mathcal{H}_{\mathrm{SOC}}
+
\mathcal{H}_{\mathrm{ex}}.
\label{eq:H_total_matrix}
\end{equation}
Here, $\mathcal{H}_{\mathrm{hop}}$ describes the multiorbital hopping, $\mathcal{H}_{\mathrm{cf}}$ is the crystal-field term, $\mathcal{H}_{\mathrm{SOC}}$ is the atomic spin-orbit coupling, and $\mathcal{H}_{\mathrm{ex}}$ is the staggered exchange field that distinguishes the altermagnetic phase from the nonmagnetic phase (see details in Supplemental Material). Compared with the AM phase, the NM phase has a larger crystal-field energy, but a smaller multiorbital hopping term and zero exchange field. In the altermagnetic phase, the enhanced multiorbital hopping gives rise to a large OHC. Moreover, the staggered exchange field reconstructs the low-energy bands and reshapes the Berry-curvature hot spots near $E_F$, thereby further enhancing the contribution from the orbital sector. 

\begin{figure}
\centering
\includegraphics[width=\linewidth]{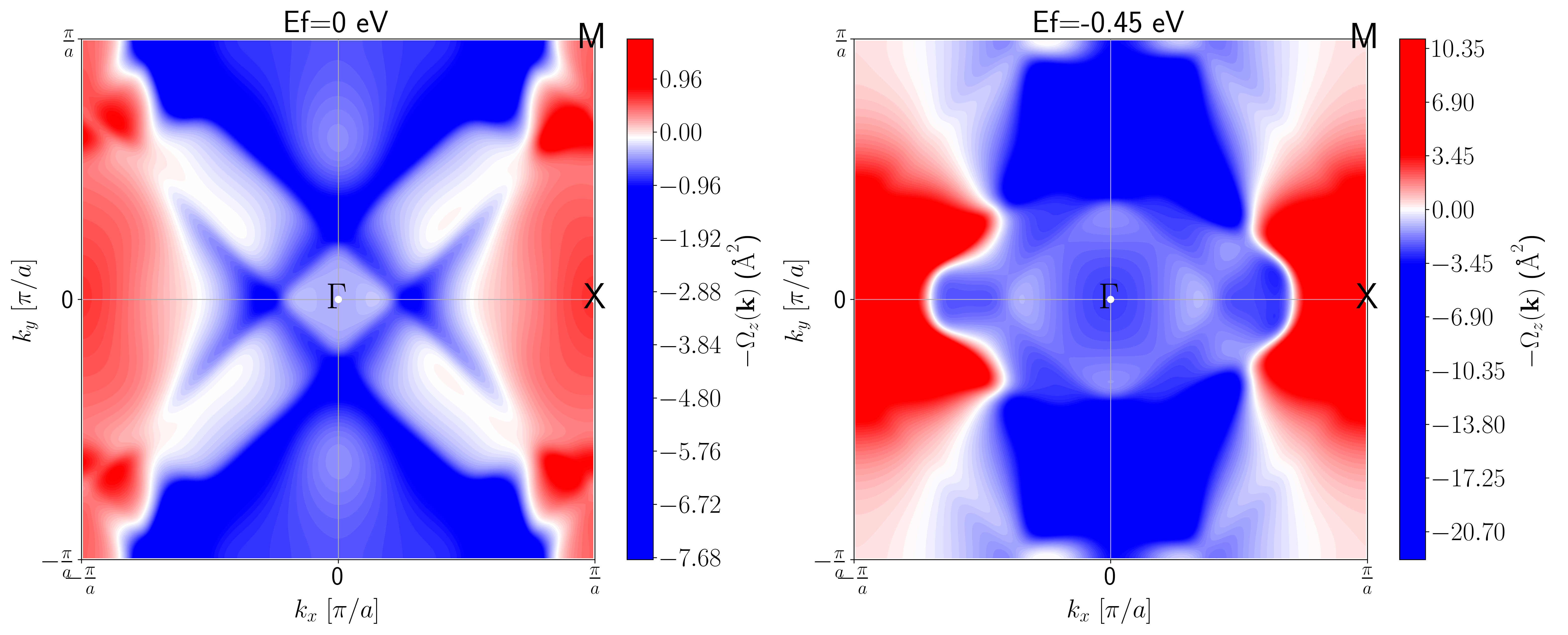}
\caption{
\justifying{
a) The calculated band structures of altermagnetic RuO$_2$ along $\Gamma-X-M-\Gamma$ without SOC. The projected orbital Berry curvature ($\Omega^{z}{(\mathbf{k})}$) onto the band structures  with SOC of b) altermagnetic and c) non-magnetic RuO$_2$. The $\Gamma-M$ path is highlighted by the shaded gray region, along which spin splitting occurs in altermagnetic RuO$_2$.}}
\label{fig:3}
\end{figure}

To find the relationship between the spin splitting and large OHE in altermagnetic RuO$_2$, we project the orbital Berry curvature onto the band structures of both AM and NM RuO$_2$ and compare them with the spin-splitting band structure of altermagnetic RuO$_2$ as shown in Fig. \ref{fig:3}. In the absence of SOC, the altermagnetic configuration exhibits clear nonrelativistic spin splitting between spin-up (red) and spin-down (blue) bands. Note that spin splitting in altermagnetic RuO$_2$ only along the $\Gamma-M$ path (and similar $Z-A$), while on other high-symmetry k-paths, symmetries still protect double degeneracy as shown in Fig. \ref{fig:3}a. By comparing spin-splitting bands (Fig. \ref{fig:3}a) with the band-projected orbital Berry curvature ($\Omega^{z}{(\mathbf{k})}$) of both AM and NM RuO$_2$ (Fig. \ref{fig:3}b and c), we can see that large orbital Berry curvatures also occur along the $\Gamma-M$ path in the altermagnetic phase, while few in the non-magnetic case. This physically explains the reason of large OHC in altermagnetic RuO$_2$. The emergence of nonrelativistic band splitting in the altermagnetic phase fundamentally changes the band structures, generating many band crossings and openings, such as those highlighted by black circles near the Fermi level along the $M-\Gamma$ path. These band changes reshape the orbital and spin Berry curvature distributions, giving rise to a dominant orbital Hall response in AM RuO$_2$. Moreover, our tight-binding model shows that AM RuO$_2$ has larger multiorbital hopping terms than the NM phase, which further contributes to its large OHC.

\begin{figure}
\centering
\includegraphics[width=0.9\linewidth]{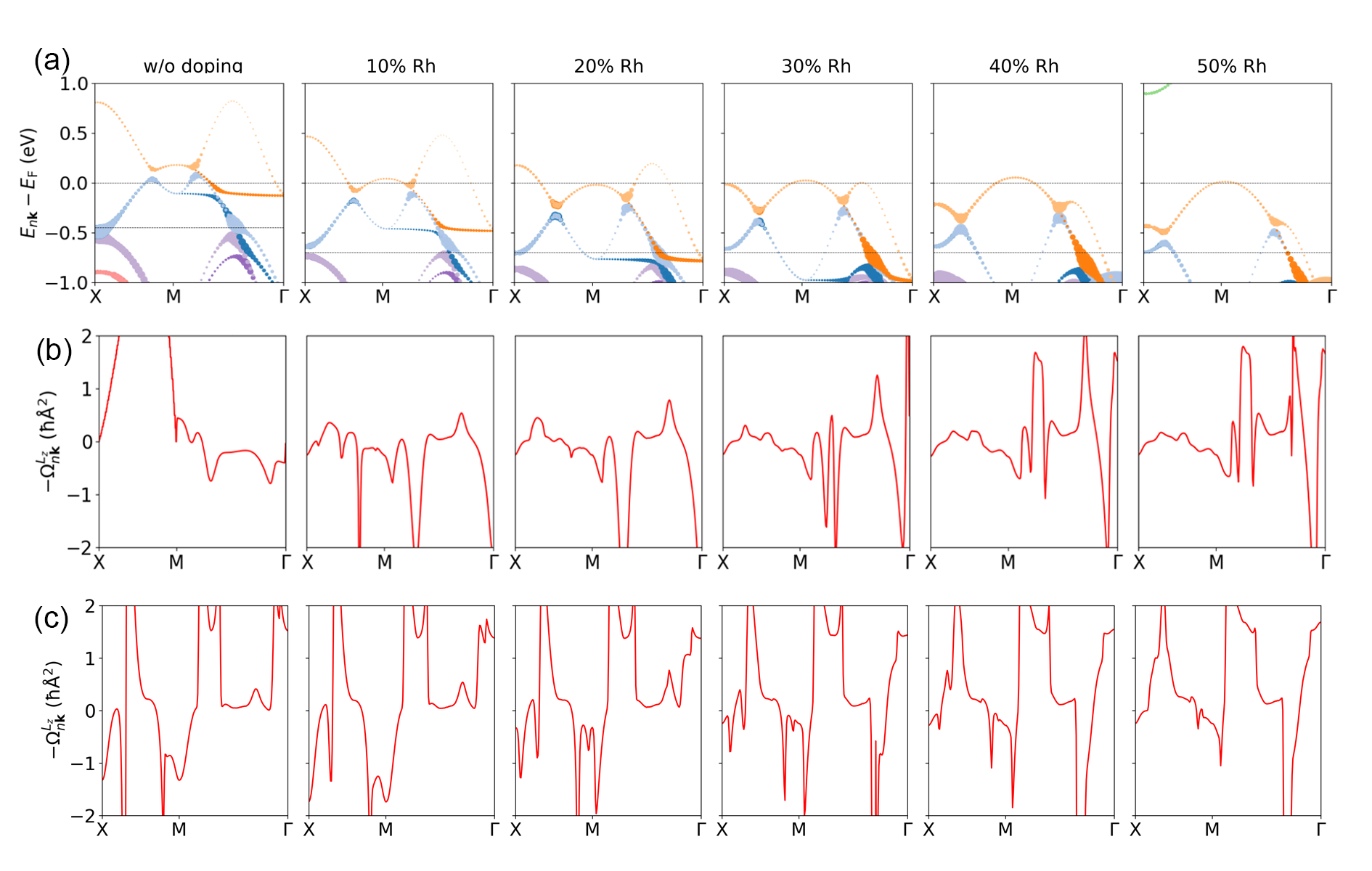}
\caption{
\justifying{
Doping-dependent orbital and spin Hall conductivities ($\sigma^{y}_{\mathrm{SHE/OHE}}$) in altermagnetic RuO$_2$ under different Rh doping concentrations. The $\textbf{S}_{(\pm)}$ and $\textbf{L}_{(\pm)}$ indicate the positive/negative spin and orbital component, respectively.}}

\label{fig:4}
\end{figure}

The anisotropic transport behavior of RuO$_2$ has been reported both experimentally\cite{liao2024separation, wang2024inverse} and theoretically\cite{PhysRevX.12.040501, yahagi2024neel}. Therefore, besides the $\sigma^{z}_{\mathrm{OHE/SHE}}$ components, we also calculate $\sigma^{y}_{\mathrm{OHE/SHE}}$ (see configurations in \textcolor{blue}{Fig.S1}) of altermagnetic RuO$_2$ as shown in Fig.~\ref{fig:4}a. As can be seen, the OHE is still much larger than SHE, but they have the same sign. Furthermore, we predict a near independence of the spin and orbital Hall effect on the orientation of the N\'eel vector, as demonstrated by similar behaviour of conductivities when the N\'eel vector is aligned along [110] direction (not shown). This implies that, in contrast to the case of the anomalous Hall effect~\cite{zhou2024crystal}, the anisotropy in the electronic structure due to antiferromagnetic ordering has a minimal influence on the SHC and OHC, explaining a level of robustness in the experimentally observed signal.

Based on the band-resolved orbital Berry curvature (Fig. \ref{fig:3}b), this orbital-dominated response is highly sensitive to band filling, enabling an efficient approach in the experiment to tune the orbital and spin Hall effects. Thus, we next investigate how to efficiently modulate OHE and SHE by electron doping with Rh which has one more electron than Ru. The evolution of band structures by Rh doping is shown in \textcolor{blue}{Fig.~S2}. Smolyanyuk et al., reported that the magnetism of RuO$_2$ can be fragile under changes in filling and composition~\cite{smolyanyuk2024fragility}. We also explicitly checked the stability of the altermagnetic state under Rh substitution and confirmed that it remains the relevant reference state in the doping range considered (see Supplemental Material).

The comparison of OHE and SHE of RuO$_2$ with varying Rh doping from 10\% upto 50\% is shown in Figs.~\ref{fig:4}b-f. We observe several interesting phenomena under electron doping. First, across different concentrations, the overall shape and magnitude of the SHE remains robust, while the OHE displays much larger variations in energy distribution and magnitude (see in \textcolor{blue}{Fig.~S3} top panel). Secondly, the magnitude of OHE gradulately increase with the increasing doping concentration. Thirdly, we find that both the relative magnitude and the sign of the OHC and SHC at $E_{\mathrm{F}}$ evolve systematically with Rh concentration. Remarkably, the orbital Hall conductivity remains negative at the Fermi level for all doping levels considered, indicating a robust orbital-dominated transverse response. In contrast, the spin Hall conductivity exhibits a sign reversal as a function of Rh concentration. It is negative at low doping levels with the same sign of OHE, and becomes positive upon increasing the Rh content beyond approximately 30\%. Lastly, around 30\% Rh, the spin Hall response can be fully suppressed at the Fermi level, while a sizable orbital Hall conductivity persists, which provides a design strategy to measure pure orbital currents in the experiment without the interference spin Hall signals. Experimentally, this orbital-dominated regime may be probed directly by MOKE in a single RuO$_2$ film, where the Kerr signal is expected to become dominated by orbital-angular-momentum accumulation once the spin Hall contribution is sufficiently suppressed. It may also be inferred indirectly in interface-sensitive heterostructures by comparing detector layers with different orbital-to-spin conversion efficiencies or SOC strengths. Overall, these results demonstrate that the coexistence of OHE and SHE in RuO$_2$ can be efficiently modulated by chemical doping in the experiment. Doping tunes both the relative strength and sign of the orbital and spin Hall effects in RuO$_2$, which gives a continuous modulation between spin-dominated, mixed spin–orbital, and purely orbital transverse transport responses, providing a rich platform for exploring novel orbitronic physical phenomena and device functionalities.

In conclusion, we have provided a physical microscopic mechanism to understand the charge--spin interconversion in RuO$_2$ by studying the orbital Hall effect and the spin Hall effect in altermagnetic and nonmagnetic phases. We show that interaction-driven altermagnetism enhances multiorbital hopping and reshapes Berry-curvature landscape, producing a dominant nonrelativistic orbital Hall response. The OHE overwhelms the relativistic spin Hall effect. It is supported by the physical orbital Berry curvature analysis that hot spots are widely distributed in momentum space and persist across a broad range of Rh doping concentrations. Under a certain doping concentration, the spin Hall response can be completely suppressed while the orbital Hall response remains finite, realizing a possible measurement of pure orbital currents in the experiment. In the non-magnetic phase, large SHE domains the nonmagnetic phase RuO$_2$. These findings naturally clarify the seemingly contradictory experimental reports on RuO$_2$, in which charge-spin interconversion signals have been attributed either to nonrelativistic altermagnetic spin splitting or to relativistic spin Hall physics. Our results show that both interpretations can be valid, depending on the underlying orbital (spin) currents, which can be further efficiently modulated by chemical doping.

\begin{acknowledgements}
L. Z and L. S. thank Chen Jiali for her discussion on results of the orbital Hall effect. L.Z is funded by the National Natural Science Foundation of China (No. 52501308), the Key Technology Research and Development Program of Shandong Province (No. 2025CXGX010406), the Natural Science Foundation of Shandong Province (No. ZR2025QC1106), and the Taishan Scholar Young Program of Shandong Province (No. tsqn202507045).  L. S. acknowledges the Ministry of Education, Singapore (Grant No. A-8001194-00-00; Grant No. A-8001872-00-00) and National Research Foundation, Singapore (Grant No. NRF-T-CRP-2025-0001). 
\end{acknowledgements}

\bibliography{ref}

\end{document}


\title{Supporting Information for:\\Coexistence and tunability of orbital and spin Hall effects in RuO$_2$}
\author{Lishu Zhang}
\email{lishu.zhang@sdu.edu.cn}
\affiliation{Key Laboratory for Liquid-Solid Structural Evolution and Processing of Materials, Ministry of Education, Shandong University, Jinan 250061, China}

\author{Mahmoud Zeer}
\affiliation{Peter Gr{\"u}nberg Institut (PGI-1) and Institute for Advanced Simulation (IAS-1), Forschungszentrum J{\"u}lich, J{\"u}lich 52428, Germany}
\affiliation{Department of Physics, RWTH Aachen University, 52056 Aachen, Germany}

\author{Dongwook Go}
\affiliation{Peter Gr{\"u}nberg Institut (PGI-1) and Institute for Advanced Simulation (IAS-1), Forschungszentrum J{\"u}lich, J{\"u}lich 52428, Germany}
\affiliation{Institute of Physics, Johannes Gutenberg University Mainz, 55099 Mainz, Germany}

\author{Theodoros Adamantopoulos}
\affiliation{Peter Gr{\"u}nberg Institut (PGI-1) and Institute for Advanced Simulation (IAS-1), Forschungszentrum J{\"u}lich, J{\"u}lich 52428, Germany}
\affiliation{Department of Physics, RWTH Aachen University, 52056 Aachen, Germany}

\author{Peter Schmitz}
\affiliation{Peter Gr{\"u}nberg Institut (PGI-1) and Institute for Advanced Simulation (IAS-1), Forschungszentrum J{\"u}lich, J{\"u}lich 52428, Germany}
\affiliation{Department of Physics, RWTH Aachen University, 52056 Aachen, Germany}

\author{Stefan Blügel}
\affiliation{Peter Gr{\"u}nberg Institut (PGI-1) and Institute for Advanced Simulation (IAS-1), Forschungszentrum J{\"u}lich, J{\"u}lich 52428, Germany}

\author{Chengwang Niu}
\affiliation{School of Physics, Shandong University, Jinan 250100, China}

\author{Yuriy Mokrousov}
\affiliation{Peter Gr{\"u}nberg Institut (PGI-1) and Institute for Advanced Simulation (IAS-1), Forschungszentrum J{\"u}lich, J{\"u}lich 52428, Germany}
\affiliation{Institute of Physics, Johannes Gutenberg University Mainz, 55099 Mainz, Germany}

\author{Shishen Yan}
\email{shishenyan@sdu.edu.cn}
\affiliation{School of Physics, Shandong University, Jinan 250100, China}

\author{Hyunsoo Yang}
\affiliation{Department of Electrical and Computer Engineering, National University of Singapore, 9 Engineering Drive 1, Singapore, 117576, Republic of Singapore}

\author{Lei Shen}
\email{shenlei@nus.edu.sg}
\affiliation{Department of Mechanical Engineering, National University of Singapore, 9 Engineering Drive 1, Singapore, 117575, Republic of Singapore}

\maketitle

\newpage

\section{Methodology}
We conducted first-principles calculations using the density functional theory (DFT) code \texttt{FLEUR}~\cite{fleurCode}, which implements the full-potential linearized augmented plane wave (FLAPW) method~\cite{Wimmer1981}. The Perdew-Burke-Ernzerhof (PBE) approximation~\cite{perdew1996generalized} was used to account for exchange and correlation effects. The effect of SOC effect was considered using the second-variation scheme in all calculations. The virtual crystal approximation (VCA) calculation, implemented within \texttt{FLEUR}, allows for the simulation of disordered alloys by averaging the potential and charge density of the constituent atoms, effectively treating the alloy as a "virtual" crystal with fractional atomic numbers. This method is particularly useful for studying systems where alloying or doping occurs but without the need for computationally challenging techniques which treat disorder more explicitly. 

To construct maximally localized Wannier functions (MLWFs) from the Bloch wave functions obtained from the self-consistent DFT calculation, we utilized the \texttt{Wannier90} package~\cite{pizzi2020wannier90}. For all systems, the MLWFs were constructed from the $p$ and $d$ orbitals of the Ru(Rh) atom, and the $p$ orbitals of the O atom. The maximum frozen window was set to be 5 eV higher than the Fermi energy for each system. A mesh of 10$\times$10$\times$12 $k$-points was employed to obtain 56 MLWFs for each system. The matrix elements of spin and orbital angular momentum operators were first evaluated in the Bloch basis and then transformed into the MLWF basis. A denser $64\times64\times64$ $k$-mesh was used for transport calculations.

The emergence of altermagnetism in RuO$_2$ is closely tied to electron-electron interactions within the Ru $4d$ orbitals. To account for the effect of electronic correlations, we utilized the GGA+$U$ method within the self-consistent DFT iteration. The Hubbard $U$ correction was employed to address the strong on-site Coulomb repulsion between the electrons in the $d$-shell of Ru(Rh). For altermagnetic phase of RuO$_2$, we selected the on-site Coulomb interaction strength $U$ to be 2.8 eV, and the intra-atomic exchange interaction strength $J$ to be 0.8 eV~\cite{zhou2024crystal}. To get non-magnetic phase of RuO$_2$, we set $U$ to be 0 eV. The calculated magnetic moment in terms of the on-site Coulomb interaction strength $U$ is presented in \textcolor{blue}{Table S1}.

The second part of this study focuses on the effects of Rh-doping on the properties of altermagnetic RuO$_2$. We employ first-principles methods to simulate the effect of replacing Ru atoms with Rh, studying the corresponding changes in the characteristic spin-splitting with the wave vector and changes in the spin and orbital conductivities. We performed self-consistent calculations  using a 10$\times$10$\times$12 Monkhorst-Pack grid in the first Brillouin zone. 

For Rh-doped RuO$_2$, full structural relaxations were performed for each doping concentration. We find that Rh substitution has a negligible effect on the lattice constants, as summarized in \textcolor{blue}{Table S2}, indicating that the overall rutile structure remains essentially intact upon doping. We found that all doped systems favor antiferromagnetic ground state under the same $U$ as that in undoped RuO$_2$. The energy difference between antiferromagnetic ground state and ferromagnetic ground state was computed to be of the order of 100\,meV. 

The orbital (spin) Hall conductivity, $\sigma_{\mathrm{OHE(SHE)}}$, is evaluated from the following expression:
\begin{equation}
\sigma_{\mathrm{OHE(SHE)}} = \frac{e}{\hbar} \sum_n \int \frac{d^3k}{(2\pi)^3} f_{nk} \Omega^z_{n,L(S)}(\mathbf{k}),
\end{equation}
where $f_{nk}$ is the Fermi-Dirac distribution function, and $\Omega^z_{n}(\mathbf{k})$ is the $z$-component of the spin (orbital) Berry curvature for band $n$ and wave vector $\mathbf{k}$. The obital Berry curvature is calculated as
\begin{equation}
\Omega^z_{n,L(S)}(\mathbf{k}) = 2\hbar^2 \sum_{m \neq n} \text{Im} \left[ \frac{\langle u_{nk} | j^z_{x,L(S)} | u_{mk} \rangle \langle u_{mk} | v_x | u_{nk} \rangle}{(E_{nk} - E_{mk} + i\eta)^2} \right],
\end{equation}
with $\langle u_{nk} | j^z_{x,L(S)} | u_{mk} \rangle$ representing the matrix elements of the conventional orbital (spin) current operator, $v_x$ is the $x$-component of the velocity operator, $E_{nk}$ is the energy of state $| u_{nk} \rangle$, and $\eta$ a positive infinitesimal number. The orbital(spin) Hall effect is thereby characterized by the integral of the orbital(spin) Berry curvature over the Brillouin zone weighted by the occupancy of each state.

\section{Altermagnetic band splitting in Rh-doped RuO$_2$ }
\begin{figure} [ht]
\centering
\includegraphics[width=0.8\textwidth]{si-figures/FigS1.pdf}
\caption{Schematic diagrams of charge-to-spin ($\sigma^{y}_{\mathrm{SHE}}$) and charge-to-orbital ($\sigma^{y}_{\mathrm{OHE}}$) current interconversion.}
\label{Fig.S1}
\end{figure}
The robustness of the altermagnetic band splitting against Rh substitution is demonstrated in \textcolor{blue}{Fig.~S2}. For all considered Rh concentrations, the electronic band structures calculated without SOC exhibit clear nonrelativistic spin splitting between spin-up and spin-down bands, confirming the persistence of the altermagnetic state upon doping. When spin-orbit coupling is included, partial hybridization between spin manifolds occurs, but the overall band topology remains largely unchanged.
In details, \textcolor{blue}{Fig.S2} shows the band evolution, where the spin-splitting is clearly visible, for different Rh concentrations. As a general feature, we observe that the SOC-induced splittings among the bands are prominently modulated by the doping, with the degree of splitting visibly increasing with Rh concentration. At higher concentrations, the impact of Rh on the band structure of undoped RuO$_2$ is drastic. With increasing Rh content, the bands become more dispersive, and local band gaps appear to be influenced by the doping level very strongly, which is especially apparent along $A-Z$. Along this path, the conduction bands move down in energy while the valence bands move up in energy with increasing doping, which results in closing of the gap around $Z$. Concerning the altermagnetic splittings, we observe that the pair of "bell"-like strongly spin-split bands positioned at the Fermi energy between $\Gamma$ and $M$ for RuO$_2$ moves down in energy without changing its shape with increasing Rh content, and so does its crossing with the stable highly dispersive pair of slightly spin-split bands which starts off at around $-$1.8\,eV at $\Gamma$ hitting the region of Fermi energy around $M$. The situation is different for the $A-Z$ path: here, the spin splitting of occupied states in the middle of the path increases by about 0.5\,eV as the Rh concentration is increased. Overall, we conclude that the incorporation of Rh introduces notable changes in the electronic structure of RuO$_2$, which could have implications for its transport properties.

\newpage
\section{Magnetic moments of RuO$_2$ as a function of $U$ }
\begin{table}[ht]
\caption{Magnetic moments of RuO$_2$ as a function of $U$ used to obtain the altermagnetic state and the nonmagnetic state.}
\label{Table.S1}
\begin{ruledtabular}
\begin{tabular}{lcccccccccccc}
\toprule
$U$ (eV) 
& 0 
& 1 
& 1.5 
& 1.7 
& 1.8 
& 1.9 
& 1.91 
& 1.92 
& 1.93 
& 1.94 
& 1.95 
& 2.0 \\
\midrule
Magnetic moment on Ru$_1$ 
& 0 
& 0 
& 0 
& 0 
& 0 
& 0 
& 0.452 
& 0.386 
& 0.455 
& 0.484 
& 0.546 
& 0.655 \\
Magnetic moment on Ru$_2$ 
& 0 
& 0 
& 0 
& 0 
& 0 
& 0 
& $-0.452$ 
& $-0.386$ 
& $-0.455$ 
& $-0.484$ 
& $-0.546$ 
& $-0.655$ \\
\bottomrule
\end{tabular}
\end{ruledtabular}
\end{table}

\clearpage
\newpage


\section{Minimal tight-binding interpretation of the phase-dependent orbital and spin Hall responses in RuO$_2$}
\label{sec:TB_model_RuO2}

To provide a transparent microscopic interpretation of the first-principles Wannier-Kubo results, we introduce here a minimal tight-binding description for the low-energy electronic states of RuO$_2$. The purpose of this model is not to quantitatively reproduce the full band structure over the entire energy window, but to identify the essential microscopic ingredients and representative low-energy energy scales responsible for the different OHC and SHC in the nonmagnetic and altermagnetic phases.

\subsection{Basis and total Hamiltonian}

The low-energy states around the Fermi level are predominantly derived from the Ru $t_{2g}$ manifold. We therefore construct a minimal model in the basis
\begin{equation}
\Psi_{\mathbf{k}}^\dagger
=
\Big(
c_{\mathbf{k}A,yz,\uparrow}^\dagger,
c_{\mathbf{k}A,zx,\uparrow}^\dagger,
c_{\mathbf{k}A,xy,\uparrow}^\dagger,
c_{\mathbf{k}A,yz,\downarrow}^\dagger,
c_{\mathbf{k}A,zx,\downarrow}^\dagger,
c_{\mathbf{k}A,xy,\downarrow}^\dagger,
c_{\mathbf{k}B,yz,\uparrow}^\dagger,
c_{\mathbf{k}B,zx,\uparrow}^\dagger,
c_{\mathbf{k}B,xy,\uparrow}^\dagger,
c_{\mathbf{k}B,yz,\downarrow}^\dagger,
c_{\mathbf{k}B,zx,\downarrow}^\dagger,
c_{\mathbf{k}B,xy,\downarrow}^\dagger
\Big),
\label{eq:basis}
\end{equation}
where $A$ and $B$ label the two Ru sublattices, $\mu,\nu\in\{yz,zx,xy\}$ denote the $t_{2g}$ orbitals, and $s,s'\in\{\uparrow,\downarrow\}$ are spin indices.

The total Hamiltonian is written as
\begin{equation}
H
=
\sum_{\mathbf{k}}
\Psi_{\mathbf{k}}^\dagger
\mathcal{H}(\mathbf{k})
\Psi_{\mathbf{k}},
\label{eq:H_total_operator}
\end{equation}
with
\begin{equation}
\mathcal{H}(\mathbf{k})
=
\mathcal{H}_{\mathrm{hop}}(\mathbf{k})
+
\mathcal{H}_{\mathrm{cf}}
+
\mathcal{H}_{\mathrm{SOC}}
+
\mathcal{H}_{\mathrm{ex}}.
\label{eq:H_total_matrix}
\end{equation}
Here, $\mathcal{H}_{\mathrm{hop}}$ describes the multiorbital hopping, $\mathcal{H}_{\mathrm{cf}}$ is the crystal-field term, $\mathcal{H}_{\mathrm{SOC}}$ is the atomic spin-orbit coupling (SOC), and $\mathcal{H}_{\mathrm{ex}}$ is the staggered exchange field that distinguishes the altermagnetic phase from the nonmagnetic phase.

\subsection{Multiorbital hopping term}

The hopping Hamiltonian is written in the general form
\begin{equation}
\mathcal{H}_{\mathrm{hop}}(\mathbf{k})
=
\sum_{\ell,\ell'=A,B}
\sum_{\mu,\nu}
\sum_{s}
\varepsilon^{\ell\ell'}_{\mu\nu}(\mathbf{k})
\,c_{\mathbf{k}\ell\mu s}^\dagger
c_{\mathbf{k}\ell'\nu s},
\label{eq:H_hop_general}
\end{equation}
where $\varepsilon^{\ell\ell'}_{\mu\nu}(\mathbf{k})$ contains the symmetry-allowed intra- and inter-sublattice hopping amplitudes between the Ru $t_{2g}$ orbitals. This term determines the multiorbital band dispersion and, importantly, the momentum-dependent orbital hybridization that underlies the orbital Hall response.

For compactness, Eq.~(\ref{eq:H_hop_general}) can be viewed as a $6\times6$ matrix in the orbital-sublattice space and is spin-independent,
\begin{equation}
\mathcal{H}_{\mathrm{hop}}(\mathbf{k})
=
h_0(\mathbf{k})\otimes \sigma_0,
\label{eq:H_hop_matrix}
\end{equation}
where $h_0(\mathbf{k})$ acts in the combined sublattice-orbital space and $\sigma_0$ is the $2\times2$ identity matrix in spin space.

In a minimal parametrization, $h_0(\mathbf{k})$ contains the dominant intra-orbital dispersions together with the interorbital hybridization terms allowed by the rutile symmetry. These hopping processes generate the orbital texture in momentum space and therefore provide the microscopic origin of a sizable OHC even in the absence of magnetic order.

\subsection{Crystal-field term}

The crystal-field contribution is taken as
\begin{equation}
\mathcal{H}_{\mathrm{cf}}
=
\sum_{\ell=A,B}
\sum_{\mu,s}
\Delta_{\mu}\,
c_{\ell\mu s}^\dagger c_{\ell\mu s},
\label{eq:H_cf_operator}
\end{equation}
where $\Delta_{\mu}$ is the onsite energy of orbital $\mu$. In matrix form,
\begin{equation}
\mathcal{H}_{\mathrm{cf}}
=
\Delta_{\mathrm{cf}}\otimes \tau_0 \otimes \sigma_0,
\label{eq:H_cf_matrix}
\end{equation}
with
\begin{equation}
\Delta_{\mathrm{cf}}
=
\mathrm{diag}(\Delta_{yz},\Delta_{zx},\Delta_{xy}).
\label{eq:Delta_cf}
\end{equation}

Here, $\tau_0$ and $\sigma_0$ are the identity matrices in sublattice and spin spaces, respectively. This term captures the crystal environment in rutile RuO$_2$ and lifts the degeneracy within the $t_{2g}$ manifold.

\subsection{Spin-orbit coupling term}

The atomic SOC on Ru is described by
\begin{equation}
\mathcal{H}_{\mathrm{SOC}}
=
\lambda
\sum_{\ell=A,B}
\mathbf{L}_{\ell}\cdot\mathbf{S}_{\ell},
\label{eq:H_SOC_operator}
\end{equation}
where $\lambda$ is the SOC strength, $\mathbf{L}=(L_x,L_y,L_z)$ is the effective orbital angular-momentum operator in the $t_{2g}$ subspace, and $\mathbf{S}=(S_x,S_y,S_z)$ is the spin-$1/2$ operator.

In the orbital basis
\(
\{|yz\rangle,|zx\rangle,|xy\rangle\},
\)
the orbital angular-momentum matrices are
\begin{equation}
L_x =
\begin{pmatrix}
0 & 0 & 0 \\
0 & 0 & i \\
0 & -i & 0
\end{pmatrix},
\qquad
L_y =
\begin{pmatrix}
0 & 0 & -i \\
0 & 0 & 0 \\
i & 0 & 0
\end{pmatrix},
\qquad
L_z =
\begin{pmatrix}
0 & i & 0 \\
-i & 0 & 0 \\
0 & 0 & 0
\end{pmatrix}.
\label{eq:L_matrices}
\end{equation}
The spin operators are
\begin{equation}
S_\alpha = \frac{\hbar}{2}\sigma_\alpha,
\qquad
\alpha=x,y,z,
\label{eq:spin_operators}
\end{equation}
where $\sigma_\alpha$ are the Pauli matrices. Equivalently, the SOC term can be written in matrix form as
\begin{equation}
\mathcal{H}_{\mathrm{SOC}}
=
\lambda
\sum_{\alpha=x,y,z}
L_\alpha \otimes \tau_0 \otimes S_\alpha.
\label{eq:H_SOC_matrix}
\end{equation}
This term couples the momentum-space orbital texture generated by multiorbital hopping to the spin sector and is therefore essential for generating the SHC from the underlying orbital structure.

\subsection{Staggered exchange field}

To describe the altermagnetic phase, we introduce a staggered exchange field,
\begin{equation}
\mathcal{H}_{\mathrm{ex}}
=
\Delta_{\mathrm{ex}}
\sum_{\ell=A,B}
\eta_\ell
\sum_{\mu}
\sum_{s,s'}
c_{\ell\mu s}^\dagger
(\hat{\mathbf{n}}\cdot\boldsymbol{\sigma})_{ss'}
c_{\ell\mu s'},
\label{eq:H_ex_operator}
\end{equation}
where $\Delta_{\mathrm{ex}}$ is the exchange energy scale, $\eta_A=+1$ and $\eta_B=-1$ distinguish the two Ru sublattices, $\hat{\mathbf{n}}$ is the direction of the local ordered moment, and $\boldsymbol{\sigma}=(\sigma_x,\sigma_y,\sigma_z)$ is the vector of Pauli matrices.

In matrix form,
\begin{equation}
\mathcal{H}_{\mathrm{ex}}
=
\Delta_{\mathrm{ex}}
\,
I_{\mathrm{orb}}
\otimes
\tau_z
\otimes
(\hat{\mathbf{n}}\cdot\boldsymbol{\sigma}),
\label{eq:H_ex_matrix}
\end{equation}
where $I_{\mathrm{orb}}$ is the identity matrix in orbital space and $\tau_z$ acts in sublattice space. In the nonmagnetic phase, $\Delta_{\mathrm{ex}}=0$, so that $\mathcal{H}_{\mathrm{ex}}$ vanishes. In the altermagnetic phase, $\Delta_{\mathrm{ex}}\neq 0$, and the staggered structure of Eq.~(\ref{eq:H_ex_matrix}) produces symmetry-allowed momentum-dependent spin splitting while maintaining zero net magnetization.

For completeness, we note that a more general model may include orbital-dependent exchange splittings and symmetry-constrained anisotropies. However, the simplified form above already captures the essential role of altermagnetic exchange in reconstructing the low-energy electronic structure relevant to the Hall responses.

\subsection{Velocity, spin-current, and orbital-current operators}

The velocity operator is defined as
\begin{equation}
\hat{v}_\alpha(\mathbf{k})
=
\frac{1}{\hbar}
\frac{\partial \mathcal{H}(\mathbf{k})}{\partial k_\alpha},
\qquad
\alpha=x,y,z.
\label{eq:velocity_operator}
\end{equation}

The spin-current operator transporting the $\gamma$ component of spin along the $\alpha$ direction is
\begin{equation}
\hat{J}_{\alpha}^{S^\gamma}
=
\frac{1}{2}
\left\{
\hat{S}^\gamma,
\hat{v}_\alpha
\right\},
\label{eq:spin_current_operator}
\end{equation}
whereas the orbital-current operator transporting the $\gamma$ component of orbital angular momentum along the $\alpha$ direction is
\begin{equation}
\hat{J}_{\alpha}^{L^\gamma}
=
\frac{1}{2}
\left\{
\hat{L}^\gamma,
\hat{v}_\alpha
\right\}.
\label{eq:orbital_current_operator}
\end{equation}
Here, $\{A,B\}=AB+BA$ denotes the anticommutator.

\subsection{Kubo formula for the Hall conductivities}

Within the intrinsic Kubo formalism, the SHC is given by
\begin{equation}
\sigma_{\alpha\beta}^{S^\gamma}
=
-\frac{e\hbar}{N_k}
\sum_{\mathbf{k}}
\sum_{n\neq m}
\frac{
f_{n\mathbf{k}}-f_{m\mathbf{k}}
}{
(\varepsilon_{n\mathbf{k}}-\varepsilon_{m\mathbf{k}})^2
}
\,\mathrm{Im}
\left[
\langle n\mathbf{k}|\hat{J}_{\alpha}^{S^\gamma}|m\mathbf{k}\rangle
\langle m\mathbf{k}|\hat{v}_{\beta}|n\mathbf{k}\rangle
\right],
\label{eq:SHC_Kubo}
\end{equation}
and the OHC is
\begin{equation}
\sigma_{\alpha\beta}^{L^\gamma}
=
-\frac{e\hbar}{N_k}
\sum_{\mathbf{k}}
\sum_{n\neq m}
\frac{
f_{n\mathbf{k}}-f_{m\mathbf{k}}
}{
(\varepsilon_{n\mathbf{k}}-\varepsilon_{m\mathbf{k}})^2
}
\,\mathrm{Im}
\left[
\langle n\mathbf{k}|\hat{J}_{\alpha}^{L^\gamma}|m\mathbf{k}\rangle
\langle m\mathbf{k}|\hat{v}_{\beta}|n\mathbf{k}\rangle
\right].
\label{eq:OHC_Kubo}
\end{equation}
Here, $|n\mathbf{k}\rangle$ and $\varepsilon_{n\mathbf{k}}$ are the eigenstates and eigenvalues of $\mathcal{H}(\mathbf{k})$, $f_{n\mathbf{k}}$ is the Fermi-Dirac distribution, and $N_k$ is the total number of $\mathbf{k}$ points used in the Brillouin-zone summation.

Equations~(\ref{eq:SHC_Kubo}) and (\ref{eq:OHC_Kubo}) show that the Hall conductivities are especially sensitive to interband matrix elements near avoided crossings and near-degenerate states, whose positions and orbital/spin characters are modified by SOC and, in the altermagnetic phase, also by the staggered exchange field (see TB parameters in Table SII).

\subsection{Representative low-energy parameters from the Wannier Hamiltonians}

To connect the above formal model to the first-principles calculations, we extract representative low-energy parameters from the Wannier Hamiltonians.

\begin{table*}[ht]
\caption{
Representative parameters in the low-energy effective tight-binding model. The hopping and onsite-like parameters are extracted from the corresponding Wannier Hamiltonians. 
}
\label{tab:model_parameters_S6_S8_S11_S15}
\begin{ruledtabular}
\begin{tabular}{lccc}
Term & Parameter & AM (SOC) & NM (SOC) \\
$H_{\mathrm{hop}}$
& $t_1$ (eV)
& 0.452
& 0.442 \\

$H_{\mathrm{hop}}$ 
& $t_2$ (eV)
& 0.276
& 0.271 \\

$H_{\mathrm{hop}}$
& $t_3$ (eV)
& 0.249
& 0.244 \\

$H_{\mathrm{cf}}$
& $\varepsilon_1$ (eV)
& 5.792
& 6.478 \\

$H_{\mathrm{cf}}$ 
& $\varepsilon_2$ (eV)
& 5.864
& 6.572 \\

$H_{\mathrm{cf}}$ 
& $\varepsilon_3$ (eV)
& 5.914
& --- \\

$H_{\mathrm{cf}}$
& $\varepsilon_4$ (eV)
& 5.966
& --- \\

$H_{\mathrm{cf}}$ 
& $\Delta_1=\varepsilon_2-\varepsilon_1$ (eV)
& 0.073
& 0.095 \\

$H_{\mathrm{cf}}$ 
& $\Delta_2=\varepsilon_3-\varepsilon_1$ (eV)
& 0.123
& --- \\

$H_{\mathrm{cf}}$
& $\Delta_3=\varepsilon_4-\varepsilon_1$ (eV)
& 0.174
& --- \\

$H_{\mathrm{SOC}}$
& $\lambda_{\mathrm{eff}}$ (eV)
& 0.008
& 0.008 \\

$H_{\mathrm{ex}}$
& $J_{\mathrm{eff}}$ (eV)
& 0.479
& 0 \\

\end{tabular}
\end{ruledtabular}
\end{table*}

Here, $t_1$, $t_2$, and $t_3$ are representative effective hopping/hybridization amplitudes extracted from the real-space Wannier Hamiltonian, i.e., $t_i=|H_{mn}(\mathbf{R}_i)|$ for selected dominant matrix elements. They represent Ru--O-mediated effective hopping channels rather than direct single-bond hoppings between isolated Ru orbitals. The onsite-like levels $\varepsilon_i$ are extracted from the $\mathbf{R}=0$ block of the Wannier Hamiltonian and characterize the low-energy level hierarchy. The splittings $\Delta_i$ are defined relative to $\varepsilon_1$, namely $\Delta_i=\varepsilon_{i+1}-\varepsilon_1$. 

As can be seen, the NM phase has a larger crystal field splitting, while the AM phase has larger orbital hopping and exchange terms. The OHE is rooted primarily in multiorbital hopping and the resulting orbital texture. The additional low-energy reconstruction in the altermagnetic phase naturally favors a more orbital-dominated response, in agreement with the first-principles OHE/SHE results shown in the main text.

\section{Group-theoretical analysis of the T-even SHC and OHC tensors in RuO$_2$}

We use the crystallographic axes $x\parallel[100]$, $y\parallel[010]$, and $z\parallel[001]$. 
For the spin/orbital Hall conductivity tensors $\sigma^{k}_{ij}$, we focus only on the
\(\mathcal{T}\)-even transverse components, i.e., the parts relevant to Hall-type responses.
The matrices below are written for each polarization channel \(k=x,y,z\), following the convention
\[
\boldsymbol{\sigma}^{k}=
\begin{pmatrix}
\sigma^{k}_{xx} & \sigma^{k}_{xy} & \sigma^{k}_{xz}\\
\sigma^{k}_{yx} & \sigma^{k}_{yy} & \sigma^{k}_{yz}\\
\sigma^{k}_{zx} & \sigma^{k}_{zy} & \sigma^{k}_{zz}
\end{pmatrix}.
\]

\paragraph{(i) Altermagnetic RuO$_2$ with SOC.}
Once spin--orbit coupling locks the magnetic order to the crystal, the \(\mathcal{T}\)-even
transverse SHC tensor takes the symmetry-allowed form
\begin{equation}
\boldsymbol{\sigma}^{x}_{\mathrm{SHC},\,\mathcal{T}\text{-even}}=
\begin{pmatrix}
0 & 0 & 0\\
0 & 0 & -a_{\mathrm{S}}\\
0 & -b_{\mathrm{S}} & 0
\end{pmatrix},
\qquad
\boldsymbol{\sigma}^{y}_{\mathrm{SHC},\,\mathcal{T}\text{-even}}=
\begin{pmatrix}
0 & 0 & a_{\mathrm{S}}\\
0 & 0 & 0\\
b_{\mathrm{S}} & 0 & 0
\end{pmatrix},
\qquad
\boldsymbol{\sigma}^{z}_{\mathrm{SHC},\,\mathcal{T}\text{-even}}=
\begin{pmatrix}
0 & c_{\mathrm{S}} & 0\\
-c_{\mathrm{S}} & 0 & 0\\
0 & 0 & 0
\end{pmatrix}.
\label{eq:SHC_Teven_SOC}
\end{equation}

The \(\mathcal{T}\)-even transverse OHC tensor has the same symmetry form,
\begin{equation}
\boldsymbol{\sigma}^{x}_{\mathrm{OHC},\,\mathcal{T}\text{-even}}=
\begin{pmatrix}
0 & 0 & 0\\
0 & 0 & -a_{\mathrm{L}}\\
0 & -b_{\mathrm{L}} & 0
\end{pmatrix},
\qquad
\boldsymbol{\sigma}^{y}_{\mathrm{OHC},\,\mathcal{T}\text{-even}}=
\begin{pmatrix}
0 & 0 & a_{\mathrm{L}}\\
0 & 0 & 0\\
b_{\mathrm{L}} & 0 & 0
\end{pmatrix},
\qquad
\boldsymbol{\sigma}^{z}_{\mathrm{OHC},\,\mathcal{T}\text{-even}}=
\begin{pmatrix}
0 & c_{\mathrm{L}} & 0\\
-c_{\mathrm{L}} & 0 & 0\\
0 & 0 & 0
\end{pmatrix}.
\label{eq:OHC_Teven_SOC}
\end{equation}

Here \(a_{\mathrm{S}}, b_{\mathrm{S}}, c_{\mathrm{S}}\) and \(a_{\mathrm{L}}, b_{\mathrm{L}}, c_{\mathrm{L}}\)
are independent symmetry-allowed coefficients in the altermagnetic phase with SOC.

\paragraph{(ii) Altermagnetic RuO$_2$ without SOC.}
In the absence of SOC, the transverse \(\mathcal{T}\)-even SHC vanishes,
\begin{equation}
\boldsymbol{\sigma}^{x}_{\mathrm{SHC},\,\mathcal{T}\text{-even}}=
\boldsymbol{\sigma}^{y}_{\mathrm{SHC},\,\mathcal{T}\text{-even}}=
\boldsymbol{\sigma}^{z}_{\mathrm{SHC},\,\mathcal{T}\text{-even}}=
\begin{pmatrix}
0 & 0 & 0\\
0 & 0 & 0\\
0 & 0 & 0
\end{pmatrix},
\label{eq:SHC_Teven_noSOC}
\end{equation}
up to possible diagonal longitudinal spin-conductivity terms, which are not relevant to the
transverse spin Hall response and are therefore omitted here.

By contrast, the \(\mathcal{T}\)-even OHC remains symmetry-allowed even without SOC, since the
orbital Hall response originates from multiorbital hopping and the corresponding orbital texture.
Its transverse tensor keeps the same symmetry pattern as in the SOC case,
\begin{equation}
\boldsymbol{\sigma}^{x}_{\mathrm{OHC},\,\mathcal{T}\text{-even}}=
\begin{pmatrix}
0 & 0 & 0\\
0 & 0 & -a_{\mathrm{L}}^{(0)}\\
0 & -b_{\mathrm{L}}^{(0)} & 0
\end{pmatrix},
\qquad
\boldsymbol{\sigma}^{y}_{\mathrm{OHC},\,\mathcal{T}\text{-even}}=
\begin{pmatrix}
0 & 0 & a_{\mathrm{L}}^{(0)}\\
0 & 0 & 0\\
b_{\mathrm{L}}^{(0)} & 0 & 0
\end{pmatrix},
\qquad
\boldsymbol{\sigma}^{z}_{\mathrm{OHC},\,\mathcal{T}\text{-even}}=
\begin{pmatrix}
0 & c_{\mathrm{L}}^{(0)} & 0\\
-c_{\mathrm{L}}^{(0)} & 0 & 0\\
0 & 0 & 0
\end{pmatrix}.
\label{eq:OHC_Teven_noSOC}
\end{equation}

\paragraph{(iii) Higher-symmetry nonmagnetic tetragonal limit.}
If the full tetragonal symmetry of nonmagnetic RuO$_2$ is restored, additional constraints reduce
the number of independent coefficients. In that limit, one recovers
\begin{equation}
b_{\mathrm{S}}=-a_{\mathrm{S}}, \qquad b_{\mathrm{L}}=-a_{\mathrm{L}},
\label{eq:tetragonal_constraints}
\end{equation}
and similarly
\begin{equation}
b_{\mathrm{L}}^{(0)}=-a_{\mathrm{L}}^{(0)}
\end{equation}
for the no-SOC OHC tensor. Thus, the nonmagnetic tetragonal phase contains fewer independent
\(\mathcal{T}\)-even Hall coefficients than the SOC-locked altermagnetic phase.

\paragraph{Physical implication.}
Equations~(\ref{eq:SHC_Teven_SOC})--(\ref{eq:OHC_Teven_noSOC}) summarize the key difference between
SHC and OHC in RuO$_2$: the \(\mathcal{T}\)-even transverse SHC requires SOC, whereas the
\(\mathcal{T}\)-even transverse OHC is already symmetry-allowed in the no-SOC limit. Therefore,
within the minimal multiorbital picture, the OHC is rooted primarily in the orbital texture generated
by hopping, while the SHC is activated mainly through SOC-induced conversion of orbital response into
spin response.


\begin{figure}[ht]
\centering
\includegraphics[width=0.6\textwidth]{si-figures/FigS2.png}
\caption{Electronic band structures of Ru$_{1-x}$Rh$_x$O$_2$ for different Rh concentrations ($x = 0.1, 0.2, 0.3, 0.4,$ and $0.5$) calculated without spin--orbit coupling (left panels) and with spin--orbit coupling (right panels). In the absence of SOC, clear nonrelativistic spin splitting between spin-up and spin-down bands is observed for all doping levels, reflecting the persistence of the altermagnetic state upon Rh substitution. When SOC is included, the bands become spin mixed while retaining the overall band topology, indicating that the altermagnetic spin splitting originates from exchange interactions rather than relativistic effects.}
\label{Fig.S2}
\end{figure}


\clearpage
\newpage
\section{Magnetic and structural characteristics of Rh-doped RuO$_2$}
\begin{table}[ht]
\caption{Magnetic properties of Rh-doped RuO$_2$. Here, $m$ denotes the local atomic magnetic moment of the transition-metal atom. The structural parameters $d_1$, $d_2$, and $d_3$ (in \AA) and the bond angle $\theta_1$ (in degrees) are defined based on the inequivalent chemical bond lengths and bond angles in the crystal structure.}
\label{Table.S2}
\begin{ruledtabular}
\begin{tabular}{cccccc}
Doping & $m$ ($\mu_B$/atom) & $d_1$ & $d_2$ & $d_3$ & $\theta_1$ \\
\hline
w/o Rh & 1.244 & 1.956 & 2.011 & 1.956 & 128.670 \\
10\% Rh & 1.159 & 1.956 & 2.010 & 1.956 & 128.673 \\
20\% Rh & 1.053 & 1.957 & 2.010 & 1.957 & 128.660 \\
30\% Rh & 0.931 & 1.958 & 2.010 & 1.958 & 128.633 \\
40\% Rh & 0.806 & 1.959 & 2.009 & 1.959 & 128.613 \\
50\% Rh & 0.712 & 1.959 & 2.009 & 1.959 & 128.608 \\
\end{tabular}
\end{ruledtabular}
\end{table}

\newpage
\begin{figure*}
\centering
\includegraphics[width=\linewidth]{si-figures/FigS3.pdf}
\caption{
\justifying{
Doping-dependent orbital and spin Hall conductivities in altermagnetic RuO$_2$. Shown are the energy-dependent OHC and SHC tensor components $\sigma^{y}_{\mathrm{OHE/SHE}}$ of RuO$_2$ for different Rh doping concentrations, ranging from the undoped case to 50\% Rh substitution. The bottom row shows enlarged views around the Fermi energy to highlight the low-energy transport behavior. The vertical dashed lines indicate the Fermi level. With increasing Rh concentration, both the magnitude and the relative sign of the OHC and SHC are continuously tuned. Notably, the orbital and spin Hall signals evolve from having the same sign at low doping to opposite signs at higher doping levels. At approximately 30\% Rh doping, the spin Hall response is completely suppressed near the Fermi energy, while a sizable orbital Hall response remains, indicating a regime dominated by pure orbital current generation. The $\textbf{S}_{(\pm)}$ and $\textbf{L}_{(\pm)}$ indicate the positive/negative spin and orbital component, respectively.}}

\label{fig:S3}
\end{figure*}

\newpage

\begin{figure*}
\centering
\includegraphics[width=1.0\textwidth]{si-figures/FigS4.pdf}
\caption{
\justifying{
Band-resolved and momentum-resolved orbital and spin Berry curvatures, as those in Fig.~\ref{fig:S3}, in magnetic Rh-doped RuO$_2$. a) The band structures of Rh-doped RuO$_2$ under different concentrations from 10\% to 50\%, where the color scale represents the magnitude of the orbital Berry curvature projected onto each band. b) The $k$-resolved orbital Berry curvature accumulated along the same high-symmetry paths, revealing pronounced orbital Berry curvature hot spots that persist and evolve with increasing Rh concentration as indicated by the black arrow. c) The corresponding $k$-resolved spin Berry curvature. Compared with the orbital counterpart, the spin Berry curvature is strongly suppressed over most of the Brillouin-zone paths as indicated by the black arrow and remains highly localized near only specific band features. The clear contrast between the orbital- and spin-resolved Berry curvatures demonstrates that the orbital Hall effect dominates the transverse response in a wide doping range.}}
\label{fig:S4}
\end{figure*}

To elucidate the physical origin of the doping-dependent orbital and spin Hall conductivities shown in Fig.~\ref{fig:S3}, we further analyze the band-resolved and momentum-resolved Berry curvature distributions in Fig.~\ref{fig:S4}. The energy-dependent Hall responses are obtained by integrating Berry curvature contributions over occupied states, and are therefore directly governed by the detailed band structure and its associated Berry curvature near the Fermi level. As shown in Fig.~\ref{fig:S4}(a), the electronic band structures of Rh-doped RuO$_2$ exhibit multiple band crossings and SOC-gaps in the vicinity of the Fermi energy, where sizable orbital Berry curvature is generated. The color-coded projections reveal that these Berry curvature contributions are predominantly orbital in character and are broadly distributed over several bands, rather than being confined to isolated band extrema. With increasing Rh concentration, the positions of these orbital-Berry-curvature-active bands shift relative to the Fermi level, leading to a substantial redistribution of orbital Berry curvature contributions to the occupied states. This behavior is further highlighted in Fig.~\ref{fig:S4}(b), which shows the accumulated $k$-resolved orbital Berry curvature along high-symmetry paths. Pronounced orbital Berry curvature hot spots persist across the entire doping range and evolve continuously with Rh concentration, reflecting the robustness of the altermagnetic band splitting and its associated orbital texture. These extended and coherent orbital Berry curvature features account for the dominant and sign-stable orbital Hall conductivity observed in Fig.~\ref{fig:S3}, including the persistence of a negative OHC at the Fermi level for all doping concentrations considered.

In contrast, the spin Berry curvature shown in Fig.~\ref{fig:S4}(c) is strongly suppressed over most of the Brillouin-zone paths and remains highly localized near specific band features. Owing to its relativistic origin, the spin Berry curvature is much more sensitive to subtle band hybridization effects induced by Rh doping. As a consequence, its integrated contribution can change sign as the relative weight of these localized hot spots shifts with doping, providing a natural explanation for the sign reversal of the spin Hall conductivity and for the near-vanishing SHC observed around 30\% Rh, where a purely orbital transverse current is realized.

Together, these results demonstrate that the strikingly different doping dependences of the orbital and spin Hall effects in RuO$_2$ originate from their distinct Berry curvature landscapes. While the orbital Hall effect is governed by robust, nonrelativistic altermagnetic band features spread over momentum space, the spin Hall effect arises from fragile, SOC-driven hot spots, leading to a tunable competition between orbital and spin angular momentum transport.

\section{Experimental guide}
\label{sec:SI_experimental_fingerprint}

To clarify how the present results may assist in identifying altermagnetism experimentally, we note that the predicted phase-dependent orbital and spin Hall responses of RuO$_2$ provide a potentially useful experimental fingerprint. In particular, based on Fig.~1, one may consider a (100) RuO$_2$/CoPt bilayer with perpendicular magnetization in the CoPt layer to distinguish between altermagnetic and nonmagnetic RuO$_2$ through the effective spin Hall angle (or conductivity). When an electric current is applied along the [010] direction of the (100) RuO$_2$ plane, the transverse angular-momentum current flows along the [100] direction and is injected into the CoPt layer, where the induced spin-orbit torque may be detected through the anomalous Hall voltage.

Assuming that the conversion efficiency from orbital current to spin current at the interface is not too small (for example, of order 10\%), our results imply qualitatively different responses for the two phases. If the (100) RuO$_2$ layer is altermagnetic, a large positive effective spin Hall conductivity, converted from the dominant orbital Hall conductivity [Fig.~1(f)], is expected. In contrast, if the RuO$_2$ layer is nonmagnetic, a clearly negative effective spin Hall conductivity, arising predominantly from the intrinsic spin Hall conductivity [Fig.~1(g)], is expected. Therefore, the sign and magnitude of the effective spin Hall signal in such a bilayer geometry may provide a practical route to distinguish the altermagnetic and nonmagnetic states of RuO$_2$.

We emphasize, however, that this transport-based distinction should be regarded as an experimentally relevant fingerprint rather than a definitive proof of altermagnetism. To unambiguously establish the altermagnetic state of RuO$_2$, a direct probe of the spin-split electronic structure, such as spin-resolved angle-resolved photoemission spectroscopy (SARPES), would still be highly desirable.

The Rh-doping-controlled evolution from spin-dominated to mixed spin--orbital and ultimately orbital-dominated regimes suggests that RuO$_2$-based materials can serve as tunable angular-momentum sources. Such tunability may be useful for optimizing charge-to-spin conversion in different heterostructures, controlling both the sign and the relative strength of the generated torque, and designing reconfigurable spin-orbitronic devices in which the dominant transverse transport channel can be adjusted by composition. In addition, for the configuration shown in Fig.~4, the SHC reverses sign when the Rh concentration exceeds approximately 30\%, whereas the OHC remains negative throughout the entire doping range considered. This contrasting behavior may provide an experimental route to identify the net contribution of the OHC to transport by systematically varying the Rh doping level.
\bibliographystyle{plainnat}
\bibliography{ref}